\documentclass[prd, aps, superscriptaddress, preprintnumbers, twocolumn, floatfix, nofootinbib]{revtex4}
\pdfoutput=1

\usepackage{amsfonts}
\usepackage{amsmath}
\usepackage{amssymb}
\usepackage{bm}
\usepackage{dcolumn}
\usepackage{graphicx}   
\usepackage[latin1]{inputenc}
\usepackage{latexsym}
\usepackage{rotating}
\usepackage{hyperref}
\usepackage{graphicx}
\usepackage{color}

\newcommand\be{\begin{equation}}
\newcommand\ba{\begin{eqnarray}}
\newcommand\ee{\end{equation}}
\newcommand\ea{\end{eqnarray}}

\newcommand{\mn}{_{\mu \nu}}

\begin{document}

\title{ Low-Energy String Theory Predicts Black Holes Hide a New Universe}

\author{Robert Brandenberger}
\email{rhb@physics.mcgill.ca}
\affiliation{Department of Physics, McGill University, Montr\'{e}al, QC, H3A 2T8, Canada}

\author{Lavinia Heisenberg}
\email{lavinia.heisenberg@phys.ethz.ch}
\affiliation{Institute for Theoretical Physics, ETH Z\"urich, Wolfgang-Pauli-Strasse 27, 8093,  Z\"urich, Switzerland}

\author{Jakob Robnik}
\email{jakob.robnik@gmail.com}
\affiliation{Institute for Theoretical Physics, ETH Z\"urich, Wolfgang-Pauli-Strasse 27, 8093,  Z\"urich, Switzerland}

\date{\today}

\begin{abstract}
We propose a construction with which to resolve the black hole singularity and enable an anisotropic cosmology to emerge from the inside of the hole. The model relies on the addition of an S-brane to the effective action which describes the geometry of space-time. This space-like defect is located inside of the horizon on a surface where the Weyl curvature reaches a limiting value. We study how metric fluctuations evolve from the outside of the black hole to the beginning of the cosmological phase to the future of the S-brane. Our setup addresses i) the black hole singularity problem, ii) the cosmological singularity problem and iii) the information loss paradox since the outgoing Hawking radiation is entangled with the state inside the black hole which becomes the new universe.

\end{abstract}

\pacs{98.80.Cq}
\maketitle

\section{Introduction}

Black hole and cosmological singularities are unavoidable if space-time is described by Einstein gravity and if the matter sources obey standard energy conditions \cite{Penrose}. Typically, these singularities correspond to subspaces in space-time where some curvature invariant diverges, and where the effective field theory description of space, time and matter based on classical Einstein gravity and classical matter fields will break down. Clearly, new physics is required both near the center of a black hole and in the very early universe. In this paper, we present a modification of the classical action which allows for a non-singular transition between a black hole and a new universe. The new physics ingredient is the addition of an S-brane to the low energy effective action. The origin of the S-brane is motivated by superstring theory: once a curvature invariant reaches the string scale, we expect towers of string states to become effectively massless \cite{swamp}, and these have to be included in the effective action which describes the space-time dynamics. S-branes have recently been used to yield a non-singular transition between a contracting and an expanding cosmology \cite{Wang1, Wang2, Wang3} (see also \cite{Kounnas}). In this setup, the S-brane arises when the background density reaches the string density. Similarly, an S-brane may arise inside the horizon of a black hole on a spacelike surface where the Weyl curvature reaches the string scale. We apply the Israel matching conditions \cite{Israel, Deruelle, Durrer} to study the space-time in the future of the S-brane, and we find either a white hole or else an expanding anisotropic cosmology, depending on the matter content to the future of the S-brane.

The idea of obtaining a black hole evaporating without a singularity goes back a long time. Early work is due to Sakharov \cite{Sakharov}, Gliner \cite{Gliner} and Bardeen \cite{Bardeen}. In particular, Sakhanov and Gliner suggested that the inside of black hole could harbor a de Sitter bubble. Note that new physics is always required in order to obtain non-singular black hole interiors \cite{Nonsing}. Various approaches to obtaining such a transition were explored, most of them involving the postulate of a matching surface presenting a local violation of the energy conditions. Examples are vacuum polarization \cite{Poisson} or quantum tunnelling \cite{Farhi}. Non-singular black holes with a de Sitter core were studied in detail by Dymnikova \cite{Dymnikova}. The effects of torsion in obtaining a cosmological space-time in the inside of a black hole was explored by Poplawski \cite{Poplawski}. In the context of string theory, it is also believed that the black hole singularity is an artefact of a low energy effective field theory description (see e.g. \cite{Veneziano} for a discussion in the context of Pre-Big-Bang cosmology). In the fuzzball picture \cite{Mathur}, string effects invalidate the effective field theory analysis all the way to the black hole horizon. In another recent approach \cite{Dvali}, the black hole interior is a non-perturbative coherent state of gravitons on top of a Minkowski space-time. In the framework of asymptotically free gravity it is also possible to obtain black holes with a non-singular interior \cite{AFree} (see also \cite{Wetterich}). For an incomplete selection of other references on black holes with non-singular interiors, see \cite{other}, and \cite{Ansoldi} for a review.

\begin{figure*}
    \begin{minipage}{0.48\textwidth}
    \includegraphics[scale = 0.34]{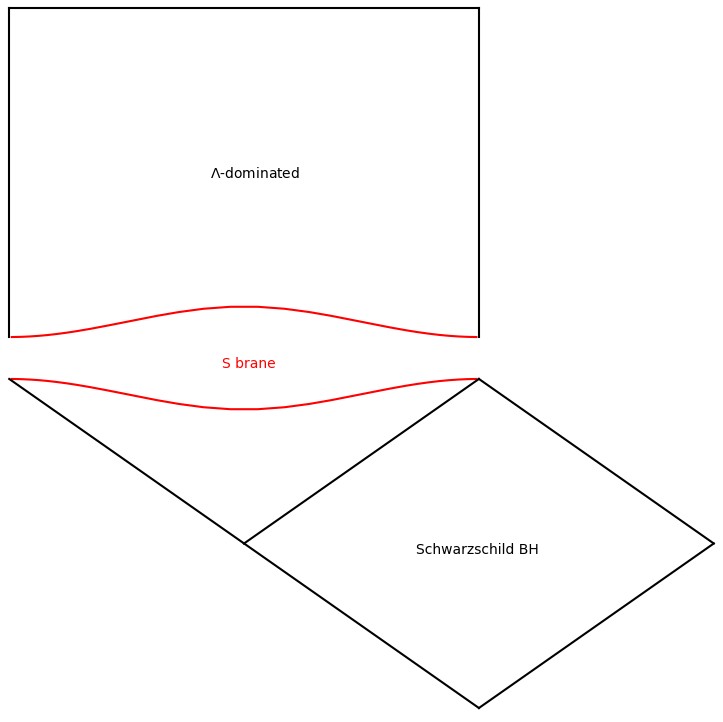}
    \end{minipage}
    \begin{minipage}{0.48\textwidth}
    \includegraphics[scale = 0.34]{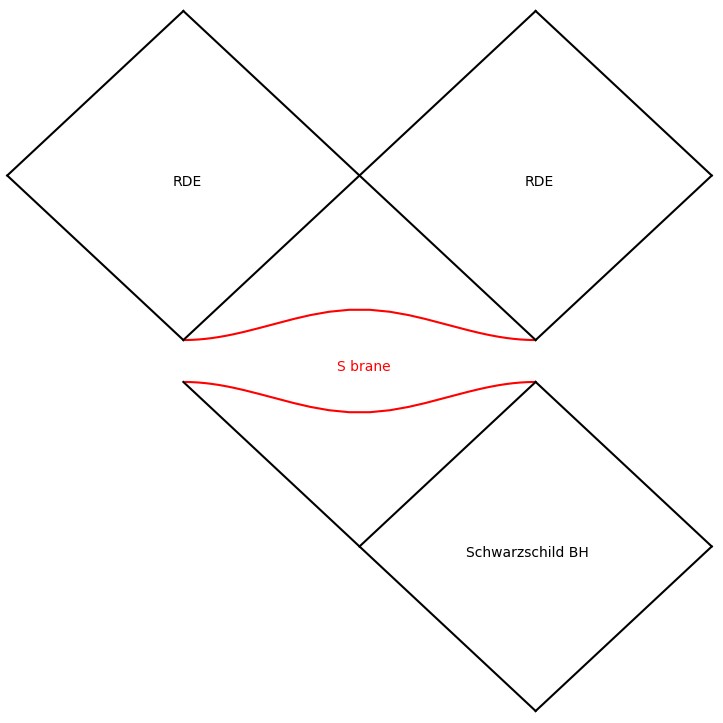}
    \end{minipage}
    \caption{Penrose diagrams of the Schwarzschild black hole matched to Bianchi universes with different matter contents to the future of the matching surface. In the left panel a $\Lambda$-dominated Bianchi universe is shown. This is asymptoticallly de Sitter space-time. In the right panel we show a radiation-dominated Bianchi universe which could be a result of the S-brane decay into photons \cite{Wang3}. Space-time becomes a time-reversed charged black hole, also known as the white hole. Similarly, a vacuum solution after the S-brane would be a Schwarzschild white hole.}
    \label{fig: conformal}
\end{figure*}

Since for a Schwarzschild black hole the singularity is spacelike, then by considering Penrose diagrams (Figure \ref{fig: conformal} it makes sense to hope that the interior of a black hole could harbor an expanding universe. In this way, the singularity resolution mechanism would simultaneously resolve both the black hole and the cosmological singularities. This idea was first extensively explored in \cite{Markov} (see also \cite{Morgan} for some other works, and \cite{Merali} for a popular science book on these ideas). In \cite{Slava}, the idea that a ``limiting curvature construction'' might lead to singularity resolution was explored. In the context of cosmology, it was shown that such a construction allows for the transition between a contracting and an expanding cosmology. The construction was applied to the case of a two-dimensional black hole in \cite{2dBH}, and in \cite{Easson} to the four-dimensional situation. However, the model suggested in \cite{Slava} has instabilities \cite{Daisuke}. A more refined construction, ``mimetic gravity'' was recently suggested \cite{mimetic}. It was shown that in this context a transition from a black hole exterior to a cosmology interior is possible \cite{Slava2}. 

Using Penrose diagrams it is also easy to see that a black hole interior can be connected to a white hole to the future of a matching surface which is placed before the black hole singularity is reached, see Figure \ref{fig: conformal}. The transition from a black hole to a white hole has recently been extensively discussed (see e.g. \cite{BH-WH} and \cite{Suddho}).
\par
Returning to the motivation for our work, we notice that as the singularity of the black hole is approached, some curvature scalars diverge and at some point the general relativity theory breaks down. In the context of string theory it is expected that a tower of string states loses mass and enters the low-energy effective action \cite{Wang1} when a critical value of curvature scalars is reached. This happens on a space-like hypersurface which we call an S-brane. The S-brane acts as an object with zero energy density and negative pressure, thus violating the null energy condition (NEC). It is named after it's cousin, the D-brane, which has zero pressure in the normal direction and negative pressure in the spatial directions. It has been shown that an S-brane with sufficient tension creates a bounce for the FLRW metric which converts a contracting universe to an expanding one \cite{Wang1}. Here, we will show that a black hole will transition to another space-time which can be interpreted either as a white hole, or as a Bianchi cosmology, depending on the matter content to the future of the S-brane. Since stringy S-branes can be shown to decay into some form of matter \cite{Wang3}), it makes sense to consider a space-time with matter to the future of the S-brane, even if the matter is unexcited in the black hole phase. The singularity is avoided and sufficient tension will force the universe to the future of the S-brane to be expanding. 

In Section \ref{sec: Background matching} we will analyze the background dynamics and derive the matching conditions due to the presence of the S-brane. In Section \ref{sec: Background evolution} we will discuss several scenarios in the future of the S-brane, depending on the dominating matter content. We then move on to the discussion of metric fluctuations about our background. We review the formalism of perturbations about a black hole background in Section \ref{sec: Perturbations introduction}, and work out the formulas for the fluctuations of the extrinsic curvature. In Section \ref{sec: Power spectrum} we study the evolution of the fluctuations on the black hole background, focusing in particular on the change in the spectral shape between the fluctuations outside of the horizon and close to the S-brane. We indeed find a change in the power index of the spectrum of fluctuations, similar to what occurs in the case of cosmology \cite{Wang2}.

\section{Background matching conditions} \label{sec: Background matching}

The low energy effective action ${\cal{A}}$ which we study is the Einstein-Hilbert action with the addition of an S-brane term:
\ba
    {\cal{A}} \, &=& \, \int dx^4 \sqrt{- \det g} \ \bigg(\frac{1}{2}\mathcal{R} +
     \mathcal{L_M} \bigg) \nonumber \\
     & & - \int dx^3 \sqrt{- \det q} \ S.
\ea
Here, $\mathcal{R}$ is the four dimensional Ricci scalar, $\mathcal{L}_M$ is the matter Lagrangian and we use units in which $c = 1$, $\hbar = 1$ and $M_{Pl}^{-2} = 8 \pi G = 1$.  $q_{\mu \nu}$ is the induced metric on the S-brane hypersurface:
\begin{equation}
    q_{\mu \nu} = g_{\mu \nu} - n_{\mu} n_{\nu} \, ,
\end{equation}
where $n$ is the normal to the surface. $S$ is a positive function representing the S-brane tension. The three-dimensional integral is taken over the space-like hypersurface (located inside of the black hole horizon) at which the critical curvature is reached. In the same way that the tension of topological defects like cosmic strings and domain walls is independent of position, we will also take $S$ to be a constant on the surface.

Note that the S-brane-term can be interpreted as an ideal-fluid matter content appearing at a space-like hypersurface. It's energy momentum tensor $S^{\mu}_{\ \nu}$ reveals that the S-brane has zero energy density $S^{0}_{\ 0}= 0$ (direction normal to the brane) and positive tension along the brane (equals negative pressure) $S^{i}_{\ j} = S \ \delta^{i}_{j}$. Hence, the S-brane violates the usual energy conditions and allows (in the context of cosmology) a transition between a contracting phase and an expanding phase, and between an exterior Schwarzschild metric and a non-singular interior. We can study the matching across the S-brane in the context of  general relativity by applying the Israel junction conditions at the S-brane.

\subsection{Israel junction conditions}

A matching hypersurface between two space-times must obey the Israel junction conditions \cite{Israel, Deruelle, Durrer}. The first condition is that the induced metric on the hypersurface is well-defined, that is, both sides of the hypersurface agree on the induced metric. The second condition is that the extrinsic curvature 
\begin{equation}
    K^{\mu}_{\ \nu} \, = \, q^{(\mu \lambda} \nabla_{\lambda} n_{\nu)},
\end{equation}
jumps by an amount given by the surface stress tensor:
\begin{equation} \label{eq: Kmatching0}
    K^{\mu}_{\ \nu} \big\rvert^{+}_{-} \, = \, S^{\mu}_{\ \nu}.
\end{equation}
We will now introduce coordinates on the black hole part of the manifold and on the part to the future of the S-brane. We will then apply the Israel junction conditions to determine initial metric of the universe to the future of the S-brane.

\subsection{Black hole}

We start with the metric of a static black hole such as the Schwarzschild solution. The line element for the black hole is
\begin{equation}\label{eq: BHmetric}
    ds^2 \, = \,  \frac{dr^2}{g(r)} - f(r) dx^2 - r^2 d\Omega^2,
\end{equation}
where $d \Omega^2 = d \vartheta^2 + \sin^2 \vartheta d \phi^2$ is the line element of the surface of a sphere, $r$ and $x$ are the Schwarzschild $r$ and $t$ coordinates which are time-like and space-like, respectively, inside the black hole event horizon. The Schwarzschild metric is a special case with 
\be
f(r) \, = \, g(r) \, = \, \frac{r_S}{r} - 1 \, ,
\ee
where $r_S$ is the Schwarzschild radius.
Close to $r = 0$, high curvatures will arise and consequently an S-brane will appear. The S-brane in these coordinates arises on a constant $r = r_0$ slice. It has the topology of $\mathbb{R} \times S^2$. The induced metric on the S-brane hypersurface is
\begin{equation}
    q_{\mu \nu} \, = \, Diag\bigg[0, -f(r_0), -r_0^2, - r_0^2 \sin^2 \vartheta \bigg]_{\mu \nu}
\end{equation}
and the extrinsic curvature is
\begin{equation}
    K^{\mu}_{\ \nu} = - Diag\bigg[ 0, \frac{\sqrt{g(r_0)} f'(r_0)}{2 f(r_0)}, \frac{\sqrt{g(r_0)}}{r_0}, \frac{\sqrt{g(r_0)}}{r_0} \bigg]^{\mu}_{\ \nu} .
\end{equation}

\subsection{Bianchi universe}

The universe after the S-brane is described by a Bianchi-type universe with a metric
\begin{equation}\label{eq: 'FRW'metric}
    ds^2  = dt^2 - a(t)^2 dx^2 - b(t)^2 d\Omega^2,
\end{equation}
where $a(t)$ and $b(t)$ are scale factors. 
In these coordinates the S-brane arises on a constant time slice $t = t_0$. The induced metric on the S-brane hypersurface and the extrinsic curvature of this surface are
\begin{equation}
    q_{\mu \nu} = Diag\bigg[ 0, -a^2, -b^2, - b^2 \sin^2 \vartheta \bigg]_{\mu \nu}
\end{equation}
and
\begin{equation}
    K^{\mu}_{\ \nu} = Diag\bigg[ 0, H_a, H_b, H_b\bigg]^{\mu}_{\ \nu} ,
\end{equation}
where the Hubble parameters are $H_a = \dot{a} / a$ and $H_b = \dot{b} / b$.

\subsection{Matching conditions}

The matching of the induced metric gives initial conditions for the scale factors
\begin{align}
    a(t_0) &= \sqrt{f(r_0)} \,,\\
    b(t_0) &= r_0 ,
\end{align}
while the jump condition of the extrinsic curvature across the S-brane hypersurface (see Equation \eqref{eq: Kmatching0}) provides initial conditions for the derivatives of $a$ and $b$:
\begin{align} \label{eq: HubbleMatching}
    H_a + \frac{\sqrt{g} f'}{2 f} &= S^{x}_{\ x} = S ,\\
    H_b + \frac{\sqrt{g}}{r_0} &= S^{\vartheta}_{\ \vartheta} = S .
\end{align}
We see that a large enough S-brane tension leads to an initially expanding universe to the future of the S-brane, in the same way that an S-brane with sufficient tension can lead to a non-singular transition between a contracting and expanding universe \cite{Wang1}.

\section{Background evolution} \label{sec: Background evolution}
We would like to determine how the scale factors $a(t)$ and $b(t)$ evolve after the S-brane transition. This will depend on the dominating matter content after the S-brane. It was shown in \cite{Wang3} that the S-brane decays in radiation, but this is not the only option. For example, it can excite a scalar matter field into a trapped metastable vacuum with a positive energy density. Therefore we will consider a collection of different options, particularly focusing on an ideal fluid matter content. 
\par

The energy-momentum tensor for an ideal fluid that we consider is
\begin{equation}
    T^{\mu}_{\ \nu} = \rho(t) Diag \bigg[1, -w_x, -w_{\Omega}, -w_{\Omega}\bigg]^{\mu}_{\ \nu},
\end{equation}
where $\rho$ is the energy density, and the equations of state for the radial and angular pressures are $P_x = w_x \rho$ and $P_{\Omega} = w_{\Omega} \rho$. The Einstein field equations give the following generalized form of the Friedmann equations:
\begin{align}
    &\frac{\dot{a}}{a} = \frac{\ddot{b}}{\dot{b}} + \frac{1 +w_x}{2} \frac{\rho b}{\dot{b}}  \label{eq: Einstein_a}\\
    &2 b\ddot{b} + \dot{b}^2 + 1 = - w_x \rho b^2 \label{eq: Einstein_b} .
\end{align}
These equations are to be solved together with the energy-momentum conservation constraint
\begin{equation}
  \nabla_\mu T^{\mu \nu} = 0 \, ,
\end{equation}
where $\nabla_{\mu}$ is the covariant derivative operator. This gives
\begin{equation} \label{eq: Einstein_rho}
    \rho \propto a^{- (1+w_x) } b^{-2 (1 + w_{\Omega})}.
\end{equation}
Note the similarity with the Friedmann result $\rho \propto a^{-3(1+w)}$. 
\par
Exact solutions for specific choices of $w_x$ and $w_{\Omega}$ have been obtained in the literature (see Section 14.3 in \cite{EinsteinExactSolutions}). Generically, a space-time after the S-brane will be a white hole. It is however possible to obtain an eternally expanding universe if the expansion is forced by matter, as we will show in the next subsection. Penrose diagrams of both cases are shown in Figure \ref{fig: conformal}. As expected, a cosmological constant and similar matter ingredients create an asymptotically de Sitter space-time, as we show in Subsection \ref{subsec: close to lambda}.

\subsection{Case $w_x =  -1$} \label{subsec: wx1}

If $w_x = -1$, the equations simplify substantially and we can even obtain power law solutions for the scale factors. This case includes radiation and a cosmological constant. 
\par
If $w_x = -1$, the energy density depends only on $b$, and Equation \eqref{eq: Einstein_b} becomes a second order equation for $b(t)$ with no explicit time dependence. Therefore it can be converted to a first order equation for $\dot{b} (b)$ with the solution:
 \begin{equation} \label{eq: firstOrderbeq}
     \dot{b}^2(b) = \sum_i \frac{\rho_0^{(i)}}{1 - 2 w_{\Omega}^{(i)}} b^{-2 w_{\Omega}^{(i)}} + \frac{C}{b} -1,
 \end{equation}
where $i$ labels different matter contents and C is an integration constant. We distinguish three cases with  
\begin{enumerate}
 \item All matter ingredients have $0 < w_{\Omega}$. For a sufficiently large $b$, the right-hand-side becomes zero and a limiting value of $b$ is reached at a finite proper time on the geodesics. $a \propto \dot{b}$ (Equation \eqref{eq: Einstein_a}) also goes to zero. This is a white hole space-time. An example of such matter content is radiation ($w_{\Omega} = 1$).
\item At least one ingredient has $w_{\Omega} < 0$ but still $-1< w_{\Omega}$. At late times we get an expanding universe with scale factors growing with a power law $b(t) \propto t ^{1/(1+w_{\Omega})}$.
\item A cosmological constant $\Lambda$ is present ($w_x = w_{\Omega} = -1$). It will create an asymptotically de Sitter space-time with a Hubble parameter $H^2 = \Lambda / 3$.
\end{enumerate}

Equation \eqref{eq: firstOrderbeq} cannot in general be integrated even if we neglect the "-1" term. However, an interesting, easily solvable case is for $w_{\Omega} = -1/4$. Then we get 
 \begin{align} \label{eq: bsolution_toy}
     b(t) &= b_0 \bigg[ 1+ 3 H_b t /2 + 3 H_b (H_a + H_b/2) t^2 \bigg]^{2/3} ,\\
     a(t) &= \frac{a_0 \ \dot{b}(t)}{b_0 H_b} , 
 \end{align}
where $H_a = H_a(t_0)$ and $H_b = H_b(t_0)$ are determined from the matching conditions. This is an expanding, anisotropic universe with scale factors behaving as power laws at late times.

\subsection{"Neighbourhood" of a cosmological constant} \label{subsec: close to lambda}
 
We have also considered some matter contents which have an equation of state close to that of a cosmological constant. We show that an ideal fluid which has an equation of state close to the cosmological constant also gives an approximately isotropic space at late times. Inflation initially smoothens the anisotropies, but they reappear once the scalar field decays.

\subsubsection{Ideal fluid}  
  
An ideal fluid can have an equation of state parameter which is close to that of a cosmological constant, that is, with both $w_x$ and $w_{\Omega}$ close to $-1$. Then, the scale factors at late times grow as power laws of time with large powers:
 \begin{equation}
     a(t) \propto t^{-1 + 1/\epsilon} \qquad b(t) \propto t^{1/\epsilon},
 \end{equation}
with 
\be
\epsilon = (1-w_x) + (1 - w_{\Omega}) /2 \, .
\ee
The universe is approximately isotropic at late times:
\begin{equation}
     \frac{H_a}{H_b} = 1- \epsilon.
 \end{equation}
 
\subsubsection{Scalar field with a potential}
 
Another possibility is to model the matter content with a slowly rolling scalar field. The matter Lagrangian is then of the form
\begin{equation}
 \mathcal{L}_M = \frac{1}{2}\partial_{\mu} \Phi \partial^{\mu} \Phi + V(\Phi) .
\end{equation}
In this case we have shown, under the slow roll approximation, that
\begin{align}
    \pm(t - t_0) &= \int \frac{d \Phi \sqrt{3 V(\Phi)}}{V'(\Phi)} ,\\
    H_b(t) &= \sqrt{\frac{V(\Phi(t))}{3}} ,\\
    H_a &= \frac{1}{H_b} (\dot{H_b} + H_b^2 + \frac{\dot{\phi}^2}{2}) .
\end{align}
In the example $V(\Phi) = V_0 \Phi^4$ we have seen that the evolution is 
\begin{equation}
    b(t) = b_0 \exp \bigg(\frac{\phi(t_0)^2}{8} \bigg(1- e^{-8 \sqrt{\frac{V_0}{3}} (t- t_0)} \bigg) \bigg) .
\end{equation}
Initially $b(t)$ grows exponentially and the universe resembles a de Sitter space-time. But when the scalar field has mostly decayed the behaviour is again as for the vacuum solution, $b(t)$ approaches a supremum value $b_0 \, \exp \phi(t_0)^2 / 8$ and $a(t)$ goes to zero. We get a white hole.

\section{Metric perturbations} \label{sec: Perturbations introduction}

We now turn to the perturbations. We will first briefly review the vector and tensor spherical decomposition and apply it to decompose a general metric perturbation. We will fix the gauge and calculate the matching conditions for all perturbations. In  Section \ref{sec: Power spectrum} we will then study the evolution of axial perturbations on the black hole side, starting from initial conditions at spatial infinity, propagate them towards the S-brane and apply the matching conditions obtained in the present section.

\subsection{Vector and tensor spherical harmonics}

A general scalar function $f \in L_2 [S^2]$ can be decomposed in the basis of spherical harmonics:
\begin{equation}
    f(\vartheta, \phi) = \sum_{l = 0}^{\infty}\sum_{m = -l}^{l} f_{l m} Y_{l m} (\vartheta, \phi).
\end{equation}
Similarly, a rank two covariant tensor on the sphere can be decomposed using a generalization to the vector and tensor spherical harmonics which we will briefly introduce here, for a full review see for example \cite{EmanueleScript}.

We will use upper case letters ($A$, $B$, ...) for the spherical indexes $\vartheta$ and $\phi$ and lower case ($a$, $b$, ...) for $r$ and $x$. We will denote the covariant derivative with respect to the 2-metric on the sphere by $\widehat{\nabla}_A$ and with respect to the full metric $g_{\mu \nu}$ by $\nabla_{\mu}$. We will from now on skip all the $l m$ indexes and sums over $l$ and $m$ for brevity. 

The metric on the sphere is
\begin{equation}
    \gamma_{A B} = \begin{bmatrix}
    1 & 0\\
    0 & \sin^2 \vartheta
    \end{bmatrix}.
\end{equation}
A Levi-Civita tensor is obtained from the Levi-Civita symbol by multiplying it by the determinant of the metric:
\begin{equation}
    \epsilon_{A B} = \sin \vartheta \begin{bmatrix}
    0 & 1 \\
    -1 & 0
    \end{bmatrix} .
\end{equation}

By taking covariant derivatives on the sphere we can construct two quantities that transform as vectors on the sphere:
\begin{equation}
    Y_A = \widehat{\nabla}_A Y \qquad S_A = \epsilon_{AB} \gamma^{BC} \widehat{\nabla}_C Y .
\end{equation}
Both are eigenvectors of the parity transformation ($\vartheta \xrightarrow[]{} \pi -\vartheta$, $\phi \xrightarrow[]{} \pi + \phi$). Their eigenvalues are $(-1)^l$ and $- (-1)^l$ respectively, which is to be compared with the $(-1)^l$ parity of the spherical harmonic $Y_{l m}$. Quantities with the same parity as the spherical harmonic $Y_{l m}$ are called polar, those with the opposite parity axial, and thus $Y_A$ and $S_A$ are polar and axial vector harmonics respectively. They will be used to decompose the mixed part of the metric perturbation $h_{a A}$. 

Polar and axial tensor spherical harmonic are, respectively:
\begin{align}
    Z_{A B} &= \widehat{\nabla}_A \widehat{\nabla}_B Y + \frac{l (l+1)}{2} Y \gamma_{A B} , \\
    S_{A B} &= \widehat{\nabla}_{(A} S_{B)} \, .
\end{align}
They will be used to decompose the angular part $h_{A B}$ of the metric perturbation.

\subsection{Spherical decomposition}

The metric perturbation
\begin{equation}
    h\mn (x) = g\mn(x) -\overline{g\mn}(x)
\end{equation}
can be decomposed in the basis of generalized spherical harmonics:
\begin{align}
h_{a b} &= Y \begin{bmatrix}
    -\overline{g_{00}} s_0 & -s_{1}\\
    -s_{1} &\overline{g_{11}} s_2 \end{bmatrix}_{a b} \\
h_{a A} &= v_a Y_A + h_a S_A \\
h_{A B} &= - \overline{g_{22}} (t_1 \gamma_{A B} Y + t_2 \widehat{\nabla}_A \widehat{\nabla}_B Y + 2 h S_{A B}),
\end{align}
where summation over $l$ and $m$ is implied. The angular dependence of the perturbations is in this way transformed to the $l$ and $m$ dependence of the coefficient functions: polar scalar ($s_0$, $s_1$, $s_2$), polar vector ($v_0$, $v_1$), polar tensor ($t_1$, $t_2$), axial vector ($h_0$, $h_1$) and axial tensor ($h$). All coefficient functions additionally depend on $r$ and $x$.

We can further take advantage of the static background ($x$-independence of the background) and decompose the $x$ dependence of the perturbations in Fourier modes. For example
\begin{equation}
    h_{l m}(r, x) = \frac{1}{2 \pi}\int h_{k l m} (r) e^{i k x} dk .
\end{equation}
We will drop the $k$ index from now on. 
\par
On the Bianchi side we replace the $r^2$ factor in the tensor perturbations by $b(t)^2$ and denote all the perturbation functions with the corresponding upper case letters.

\subsection{Gauge choice}

We will now examine how the metric perturbations transform uunder a coordinate change:
\begin{equation}
    \widetilde{x^{\mu}} = x^{\mu} + \xi^{\mu}.
\end{equation}
A general perturbative coordinate change $\xi^{\mu}$ can be decomposed into scalar and vector parts:
\begin{equation}
    \xi =  Y \Xi^{a} \partial_a + (\Xi^P Y^{A} + \Xi^A S^{A}) \partial_A \, ,
\end{equation}
where $\Xi^0$, $\Xi^1$, $\Xi^P$ and $\Xi^A$ are functions of $r$, $x$, $l$ and $m$.

The metric in the new coordinates is
\begin{equation}
    \widetilde{g\mn}(\widetilde{x}) = \overline{g\mn}(\widetilde{x}) + h\mn - [\mathcal{L}_{\xi} \ \overline{g} ]\mn \equiv \overline{g\mn}(\widetilde{x}) + \widetilde{h\mn} \, .
\end{equation}
The Lie derivative term is
\begin{equation}
    [\mathcal{L}_{\xi} \ \overline{g} ]\mn = \xi^{\alpha} \partial_{\alpha} \overline{g_{\mu \nu}} + 2\overline{g_{\alpha (\mu}} \partial_{\nu)} \xi^{\alpha}  = \overline{g_{\alpha (\mu}} \nabla_{\nu)} \xi^{\alpha}.
\end{equation}
It can be calculated most conveniently by spelling out the covariant derivative and recognizing the $\widehat{\nabla}$ terms. We obtain the transformation rules for the coefficient functions (written for the Bianchi metric):
\begin{align}
    \widetilde{S_0} &= S_0 - 2 \dot{\Xi}^0 \\ \nonumber
    \widetilde{S_1} &= S_1 + a^2 \dot{\Xi}^1 - \partial_x \Xi^0 \\ \nonumber
    \widetilde{S_2} &= S_2 + 2 (H_a \Xi^0 + \partial_x \Xi^1) \\\nonumber
    \widetilde{V_0} &= V_0 + \Xi^0 - (2 H_b + b^2 \partial_t) \Xi^P \\ \nonumber
    \widetilde{V_1} &= V_1 - a^2 \Xi^1 - b^2 \dot{\Xi}^P\\ \nonumber
    \widetilde{T_1} &= T_1 - H_b \Xi^0 \\ \nonumber
    \widetilde{T_2} &= T_2 - 2 \Xi^P \\ \nonumber
    \widetilde{H_0} &= H_0 - (2 H_b + b^2 \partial_t) \Xi^A \\ \nonumber
    \widetilde{H_1} &= H_1 - b^2 \dot{\Xi}^A \\ \nonumber
    \widetilde{H} &= H - \Xi^A .
\end{align}

We will fix a gauge. Setting $\widetilde{T_2} = 0$ and $\widetilde{H} = 0$ will completely fix $\Xi^P$ and $\Xi^A$ respectively. Then $\widetilde{V_1} = 0$ will fix $\Xi^1$. A Regge-Wheeler gauge would be to further set $\widetilde{V_0} = 0$ by fixing $\Xi^0$, we will discuss this in the next subsection. For the dipole case there exists no tensor spherical harmonic function and hence the parametrization of the metric perturbation using two spherical harmonic coefficients is redundant. For $l=0$ the Regge Wheeler gauge does not fix the gauge entirely (see \cite{Kobayashi:2014wsa, EmanueleScript} for a pedagogical exposure). We will ignore them in this work.

\subsection{S-brane position perturbation}

The S-brane arises on a hypersurface of constant Weyl or Kretschman scalar. Let $q(\boldsymbol{x})$ be a scalar which is constant on the S-brane. It is slighly perturbed in the presence of perturbations and it thus may no longer depend on $x^0$ only:
\begin{equation}
    q(\textbf{x}) = q_0 (x^0) + \delta q(\textbf{x}) .
\end{equation}
Therefore an S-brane is no longer a constant $x^0$ surface. We will therefore use a different time-like coordinate
\begin{equation}
    \widetilde{x^0} = x^0 + \Xi^0(\textbf{x})
\end{equation}
such that the S-brane is a constant $\widetilde{x^0}$ hypersurface. In other words, we want:
\begin{align}
    q(\widetilde{\textbf{x}}) &= q_0 (\widetilde{x^0}-\Xi^0) + \delta q (\widetilde{\textbf{x}}) \\ \nonumber
    &\approx q_0(\widetilde{x^0}) - q_0'(\widetilde{x^0}) \Xi^0 + \delta q
\end{align}
to be a function of $\widetilde{x^0}$ only. Therefore we must choose (as in \cite{Deruelle}):
\begin{equation} \label{eq: Tchoice}
    \Xi^0(\textbf{x}) = \frac{\delta q(\textbf{x})}{q_0'(x_0)} \, ,
\end{equation}
at least in a neighbourhood of the S-brane. This in principle exhausts all the remaining gauge freedom. What concerns the axial perturbations, this gauge choice is the same as the Regge-Wheeler gauge, because axial perturbations do not transform with $\Xi^0$, but polar vector modes $V_0$ cannot be set to zero because we have already chosen $\Xi^0$. 

\subsection{Induced metric}

We will first match the induced metric. The normal to the S-brane hypersurface is perturbed
\begin{equation}
    n = \sqrt{g_{00}} dx^0 + \sqrt{g^{00}} dx^1 + \sqrt{g^{00}} h_{0 A} dx^A,
\end{equation}
such that it remains orthogonal to all three spatial covectors $dx^i$. As expected, we find that the induced metric of each mode is the mode's $h_{\mu \nu}$ with $0 \mu$ and $\mu 0$ components set to 0. Matching the induced metric yields:
\begin{align}
    S_{2}(r_0, x) &= s_{2}(t_0, x) \\ \nonumber
    T_{1}(r_0, x) &= t_{1}(t_0, x) \\ \nonumber
    H_{1}(r_0, x) &= h_{1}(t_0, x)
\end{align}
for all $x \in \mathbb{R}$.

\subsection{Extrinsic curvature}

The extrinsic curvature matching conditions are most conveniently decomposed if expressed in terms of a purely covariant extrinsic curvature tensor. Neglecting possible S-brane tension perturbations, the matching condition is:
\begin{equation} \label{eq: perturbationMatchingCondition}
    \delta K_{\mu \nu}\big\rvert^{+}_{-}= h_{\mu \alpha} S^{\alpha}_{\ \nu} \, .
\end{equation}
The extrinsic curvature perturbation can be decomposed in the spherical harmonic basis as was the metric. Due to orthogonality, matching is only between modes with the same $l$ and $m$ and between modes of the same parity. The $xA$ component of Equations \eqref{eq: perturbationMatchingCondition} gives the axial matching condition

\begin{align}
    &\dot{H_1} + \sqrt{f} h_1' = H_b H_1 + \frac{\sqrt{f}}{r} h_1 . \label{eq: matchingaxialvector}
\end{align}

From the $xx$ and $\theta \theta$ components we get the polar matching conditions
\begin{align}
    \dot{S_2} + \sqrt{f} {s_2}'&= H_a S_0 + \frac{f'}{2\sqrt{f}} s_0\\ \nonumber
    \dot{T_1} + \sqrt{f} {t_1}'&= H_b S_0 + \frac{\sqrt{f}}{r} s_0 \, .
\end{align}
We will use the axial matching conditions in the next Section to obtain the axial power spectrum after the S-brane transition.

\section{Axial perturbations power spectrum} \label{sec: Power spectrum}

Now we explore the evolution of the perturbations. We are given some initial data at $r = \infty$ in the black hole space-time and would like to propagate the perturbations to the S-brane and apply the matching conditions to get the initial power spectrum at the onset of the cosmological phase. We will focus on the axial perturbations for simplicity.

In Subsection \ref{subsec: effective potential} we will first derive the governing equation for the evolution of the fluctuations from the perturbed Einstein field equations. This is a second order ordinary differential equation for the perturbation variable. Solving it will give the perturbations just before the S-brane, given the initial condition. Initial condition will be discussed in Subsection \ref{subsec: initial condition}. It is well known that the region exterior to the event horizon transmits waves with high $k$ and suppresses the low $k$ modes \cite{scattering}. To understand the transfer through the entire black hole part of the space-time we will further be interested in the transfer through the region interior to the event horizon. We will present three methods for calculating this transfer. First, in Subsection \ref{subsec: order of magnitude} we will give an order of magnitude estimate. In Subsection \ref{subsec: region solution} we will obtain an analytical result in the low $k$ limit. We will then complement this analysis with a numerical calculation, using Leaver's series expansion (Subsection \ref{subsec: Leaver}).

In Subsection \ref{subsec: Sbrane matching} we will finally use the resulting power spectrum just before the S-brane and apply the matching conditions from the previous Section \ref{sec: Perturbations introduction} to find the power spectrum at the onset of the Bianchi cosmology.

\subsection{Effective potential} \label{subsec: effective potential}

Assuming there is no axial energy-momentum tensor perturbation, the $R_{\vartheta \phi}$ component of the Ricci tensor perturbation shows us that $h_1$ and $H_1$ are not dynamical as they are completely fixed by $h_0$ and $H_0$:
\begin{equation}
    h_1= \frac{f (f h_0)'}{i k} \qquad H_1 =  \frac{a (a H_0)^{\boldsymbol{\cdot}}}{i k} \, .
\end{equation}
Combining with $R_{t \phi}$ we can formulate the evolution of $h_0$ as a one-dimensional Schr\"{o}dinger equation:
\begin{equation} \label{eq: schrodinger}
    \psi'' + V \psi = 0 \, ,
\end{equation}
where $\psi = h_0 \ f^{3/2} /r$ for the black hole part of the space-time and $\psi = H_0 \ a^{3/2} b^{-1}$ for the Bianchi part of the space-time. The effective potentials are
\begin{align}
    V^{(BH)}(r) &= \frac{L}{f r^2}+ \frac{k^2}{f^2} + \frac{3 f'}{f r} + \bigg(\frac{f'}{2 f}\bigg)^2 - \frac{f''}{2 f} \\ 
    V^{(Bian)}(t) &= \frac{L}{b^2}+ \frac{k^2}{a^2} + \\ \nonumber
    &+ 3 H_b (H_a + H_b) + 3 \dot{H_b} - \dot{H_a}/2 \, ,
\end{align}
where $L = l (l+1)$ for short. 
\par
Specifically, for the Schwarzschild black hole
\begin{equation} \label{eq: Schwarzschild potential}
    V(R) = \frac{K^2 R^ 4+  L R (1 - R) + 4 R - \frac{15}{4}}{ R^2 (1 - R)^2} \, ,
\end{equation}
where we have changed to a dimensionless radial coordinate $R = r / r_S$ and a dimensionless $K = k \ r_S$.
We call the first term the $K$-term, the second term the $L$-term and the rest the geometry-term. It is straightforward to analyze the behaviour of $\psi$ close to the singularity ($R \xrightarrow[]{} 0$), close to the event horizon ($R \sim 1$) and at infinity ($R \xrightarrow[]{} \infty$):
\begin{align} \label{eq: regionsolutions}
    \psi(R \xrightarrow[]{} 0) &= C_{\infty} \ R^{-3/2} + C_{0} \  R^{5/2} \\ \nonumber
    \psi(R \sim 1) &= \vert 1-R \vert^{1/2} \\ \nonumber& \bigg( C_+^{H} \ e^{i K \log \vert 1-R \vert} + C_-^{H} \ e^{-i K \log \vert 1-R \vert} \bigg)  \\ \nonumber
    \psi(R \xrightarrow[]{} \infty) &= C_+ e^{i K R} + C_- e^{-i K R}, \
\end{align}
where the various $C$ are integration constants. 
\par
The quantity of interest is the transfer function
\begin{equation} \label{eq: transfer function}
    \mathcal{T} = \frac{\vert C_{\infty} \vert}{\vert C_{-}^{H} \vert},
\end{equation}
which will allow us to compute the power spectrum just before the S-brane, given the initial conditions. 

Some comments are in place. Close to the singularity the $R^{-3/2}$ term prevails and the perturbations diverge, faster than the background metric. Nevertheless, we will assume that $C_{\infty}$ is small enough, so the perturbations are still linear at the S-brane. The $R^{-3/2}$ term is dominant, and as we will see, $C_{\infty}$ will determine the power spectrum after the S-brane. Second, note that $\psi$ does not appear to be smooth at the horizon. This is a coordinate artefact which is resolved by using for example tortoise coordinates as in \cite{EmanueleScript}. However, the smoothness requirement translates to the condition that only the incoming waves are present, so $C_+^{H}$ has to be zero.

Several methods for solving the Equation \eqref{eq: Schwarzschild potential} have been proposed in the literature, the WKB approximation \cite{WKB}, the Leaver series solution \cite{Leaver}, the asymptotic iteration method \cite{asymptoticIterationMethod} and others. We will first solve the differential equation by dividing the space-time into regions where the potential is simple, determine the scaling of the perturbations in each region separately and glue the solutions together. This will give us physical insight into the behaviour of the perturbations. However, this solution is very rough so we will complement our analysis with an analytical solution in the low $K$ limit and with the Leaver series solution to get accurate results. 

\subsection{Initial condition} \label{subsec: initial condition}
We define an initial power spectrum, far away from the black hole as (see Equation \eqref{eq: regionsolutions}):
\begin{equation}
    P_i(K, L) = \vert C_+ \vert^2 + \vert C_- \vert^2 .
\end{equation}
One simple choice is that this initial power spectrum originates from the quantum fluctuations in the asymptotically flat region. Quantized $\psi$ is a massless scalar field. Power spectrum is a Fourier transform of it's correlator, yielding:
\begin{equation}
    P_i(R) \propto \bigg( K^2 + \frac{L}{R^2} \bigg)^{-1/2} \approx K^{-1} \, .
\end{equation}
for large $R$. 

For large values of $K$ it is known that the region exterior to the event horizon perfectly transmits the perturbations \cite{scattering}, so the power spectrum at the horizon will be the same as the power spectrum at $r = \infty$. Small $K$ modes on the other hand are suppressed outside of the event horizon. We will in the following subsections show that they are also suppressed inside the horizon implying that they will also be suppressed on the Bianchi side of the space-time.

\subsection{Order of magnitude estimate} \label{subsec: order of magnitude}

We divide the interior of the event horizon into regions such that we approximately understand the scaling of $\psi$ in each region. By combining these scalings we estimate the transfer function in the $K >> L$ limit and in the $K << L$ limit.

\subsubsection{High $K$}

For $K^2 >> L$ the interior of the event horizon can approximately be divided in two regions. Close to the singularity the geometry-term in the effective potential dominates (Equation \eqref{eq: Schwarzschild potential}), while for larger $R$ the $K$-term dominates. A regime-change occurs approximately at the radius $R_1$ at which both terms are equally important:
\begin{equation}
    R_1 = \bigg( \frac{15}{4 K^2} \bigg)^{1/4}.
\end{equation}
The dominant solution for $\psi$ scales as $R^{-3/2}$ in the geometry-dominated region (Equation \eqref{eq: regionsolutions}) and as $R^{-1/2}$ in the $K$-dominated region at low $R$ (this can be seen by using the WKB approximation which is a good approximation for high $K$). Requiring that both solutions agree on the value of $\psi$ at $R_1$ 
\begin{equation}
    C_{\infty} R_{1}^{-3/2} = C_{-}^{H} R_1^{-1/2},
\end{equation}
yields an order of magnitude estimate for the transfer function (Equation \eqref{eq: transfer function}):
\begin{equation}
 \mathcal{T} \sim K^{-1/2}.
\end{equation}
For high $K$, the $L$ dependence of the power spectrum is preserved when passing through the interior of the event horizon, while higher $K$ modes are suppressed relative to the lower $K$ modes.

\subsubsection{Low $K$}

For low $K$, the interior of the event horizon can approximately be divided into three regions
\begin{itemize}
    \item Region I = $(R_0, \ R_{12})$ \textit{(close to the singularity)}: the geometry-term dominates,
    \item Region II = $(R_{12}, \ R_{23})$: the $L$-term dominates, except for very low $L$,
    \item Region III = $(R_{23}, \ R_{34})$ \textit{(close to the event horizon)}: the $K$-term and geometry-term dominate.
\end{itemize}
Note that $R_0 = r_0/r_S$ is the position of the S-brane. Other radii are defined as radii at which the suitable terms are equally important, e.g. $R_{12}$ is a radius at which the geometry-term and the $L$-term become equally important. They are solutions of polynomial equations and are for $l \geq 3$ and $K << 1$, up to a good approximation, equal to:
\begin{align}
    R_{12} & = \frac{15}{4 L} \, ,\\ \nonumber
    R_{23} &= 1 - \frac{1 + 4 K^2}{4 L} \, .
\end{align}
Using the scalings of $\psi$ in Regions I and III (Equation \eqref{eq: regionsolutions}), and assuming that $\psi$ is of approximately constant amplitude in Region II yields
\begin{equation}
    C_{\infty} R_{12}^{-3/2} \approx C_{-}^{H} (1 - R_{23})^{1/2} ,
\end{equation}
and one can estimate the transfer function:
\be
\mathcal{T} = \frac{\sqrt{1 + 4K^2} \, 15^{3/2} }{(4 L)^2} \, .
\ee
At low $K$, the lower $L$ and higher $K$ modes are enhanced relative to the higher $L$ and lower $K$ modes. In the following two subsections we turn to a progressively improved analysis, but the qualitative behaviour is correctly predicted by the above estimates.

\subsection{Region solution} \label{subsec: region solution}

We will here calculate the transfer function for low $K$ analytically, with very accurate results. Close to the event horizon the $L$-term is negligible and $\psi$ behaves as (see Equation \eqref{eq: regionsolutions}):
\begin{equation} \label{eq: psiH}
    \psi_H = C_{-}^{H} \, (1-R)^{-i K + 1/2} \, .
\end{equation}
As $R$ decreases the $L$-term becomes ever more important while the $K$-term becomes ever more negligible. A regime-change occurs approximately at the radius $R_1$ at which both terms are equally important: $L R_1 (1 - R_1) =  K R_1^4$. To lowest order in a small $K$ expansion we get 
\begin{equation}
    R_1 = 1 - \frac{K^2}{L}.
\end{equation}
$1 - R_1$ is of the order of $K^2$ and will turn out to be negligible for the lowest order estimate of the transfer function.

The governing Equation \eqref{eq: schrodinger} is solvable by elementary functions in the region (0, $R_1$), where the $K$-term is neglected. We call the basis functions for the space of solutions $u_l(R)$ and $v_l(R)$. We can chose $u_l(R)$ such that it is the extension of the $R^{5/2}$ solution near the singularity (see Equation \eqref{eq: regionsolutions}) to the entire region (0, $R_1$). Similarly, we can chose $v_l(R)$ as an extension of the $R^{-3/2}$ solution.

$u_l(R)$ is of the form:
\begin{equation} \label{eq: ul}
    u_l(R) = R^{5/2} \sqrt{1 - R} \ p_{l}(R) \, ,
\end{equation}
where $p_l(R)$ is a polynomial of order $l-2$ in $R$:
\begin{equation}
    p_l(R) = \sum_{n = 0}^{l -2} a_n R^n.
\end{equation}
A recurrence relation for the coefficients of the polynomial is:
\begin{equation}
    \frac{a_{n+1}}{a_n} = \frac{(n+3)(n+2) - l (l+1)}{(n+1)(n+5)},
\end{equation}
By requiring that $u$ is an extension of the $R^{5/2}$ solution we fixed $p_l(0) = a_0 = 1$. For example, for $l = 2, 3, 4$ one has:
\begin{align}
    p_2(R) &= 1 \\ \nonumber
    p_3(R) &= 1- (6/5) R \\ \nonumber
    p_4(R) &= 1 - (14/5) R + (28/15) R^2.
\end{align}
Note that for $K = 0$, $u_l$ is the solution in the entire region $(R_0, \, 1)$ and satisfies the initial condition at the horizon. Therefore $C_{\infty} = 0$. For $K \neq 0$ we will glue the solution in the region close to the horizon with the solution in the rest of the interior of the black hole, by requiring that $\psi$ and $\psi'$ are continuous at $R_1$:
\begin{align}
    C_{0} u_l(R_1) + C_{\infty} v_l(R_1) = \psi_H(R_1) \\ \nonumber
    C_{0} u_l'(R_1) + C_{\infty} v_l'(R_1) = \psi_H'(R_1) .
\end{align}
Solving for $C_{\infty}$ gives
\begin{equation}
    C_{\infty} = \frac{\psi_H' u - \psi_H u'}{\mathcal{W}},
\end{equation}
where the Wronskian determinant is $\mathcal{W} = u v' - u' v$. A standard result in the theory of differential equations is that the Wronskian determinant is independent of $R$ if the governing equation has no term proportional to $\psi'$, as is the case in the Equation \eqref{eq: schrodinger}. The Wronskian determinant can therefore be conveniently calculated in the vicinity of the singularity and equals $\mathcal{W} = - 4$. 

Substituting $\psi_H$ from Equation \eqref{eq: psiH} and $u_l$ from Equation \eqref{eq: ul}, neglecting terms beyond linear order in the small $K$ expansion, and dividing by $C_0$ gives the transfer function (Equation \eqref{eq: transfer function}):
\begin{equation}
   \mathcal{T} = \frac{\vert p_l(R = 1)\vert \ K}{4}. 
\end{equation}
The transfer function is linear in $K$. The proportionality constant $\vert p_l(1) \vert  \, / 4 = \vert \sum_{n = 0}^{l-2} a_n \vert \, / \, 4$ as a function of $l$ is shown in Table \ref{table}. It decays approximately as $L^{-2} \approx l^{-4}$.

\begin{table*}
\caption{The transfer function for low $K$ is proportional to $K$. The constant of proportionality decays as $L^{-2}$ for large $L$. We report the constant of proportionality multiplied by $L^2$ and rounded to two decimal places. The numerical solution, using the Leaver series expansion of Subsection \ref{subsec: Leaver}, gives the exact same results (within the numerical accuracy) as the analytical result of Subsection \ref{subsec: region solution}} \label{table}
\begin{tabular}{|c|c|c|c| c| c| c| c| c| c|} 
 \hline
 l & 2 & 3 & 4 & 5& 6& 7 & 8 & 9& 10\\ \hline
 $\mathcal{T} L^2 / K$ & 9 & 7.20 & 6.67 & 6.43 &  6.30& 6.22& 6.17& 6.14& 6.11 \\
 \hline
\end{tabular}
\end{table*}

\subsection{Leaver's series solution} \label{subsec: Leaver}

\begin{figure*}
    \centering
    \includegraphics[scale = 0.35]{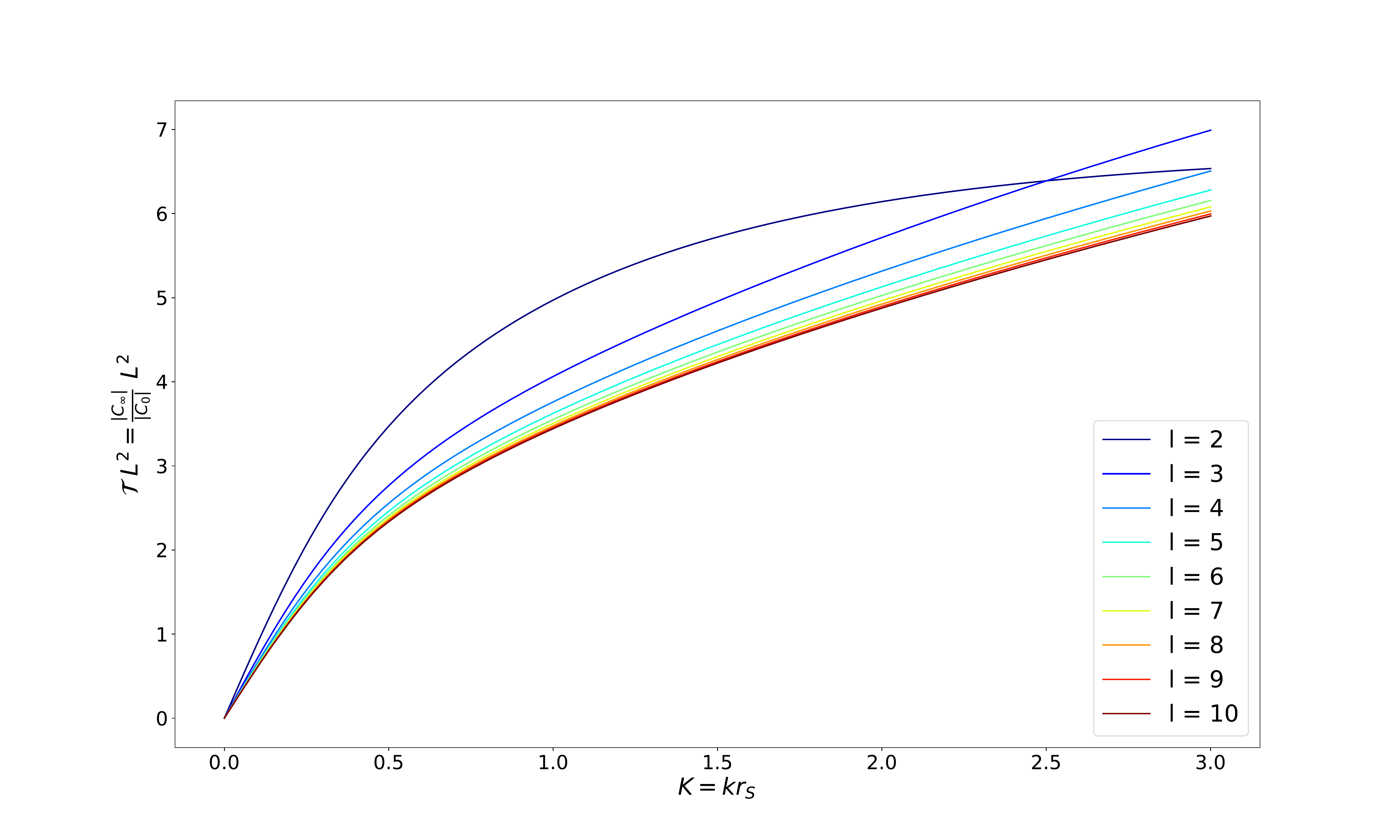}
    \caption{The axial power spectrum transfer function for the interior of the Schwarzschild black hole event horizon is shown, calculated using the Leaver series expansion (Subsection \ref{subsec: Leaver}). The simplified analysis suggests it should scale as $L^{-2}$, except for low $l$. The plot confirms this and shows the low-$l$ correction. The transfer function at low $k$ is linear as predicted by the analytic solution of Subsection \ref{subsec: region solution}.}
    \label{fig: LeaverCinf}
\end{figure*}

We now solve the evolution Equation \eqref{eq: schrodinger} with the Leaver series expansion \cite{Leaver}, to confirm the approximation in the previous subsection and to go beyond the lowest order in $K$. The Leaver series expansion is a very accurate method and was used extensively for calculating black hole quasi-normal mode (QNM) frequencies \cite{EmanueleScript}. The idea is to use an asymptotic series expansion with an ansatz which has the correct behaviour at the borders between the regions. The ansatz for the interior of the event horizon is
\begin{equation} \label{eq: Leaver ansatz}
    \psi(R) = R^{-3/2} (1-R)^{-i K + 1/2} \sum_{n = 0}^{\infty} b_n (1-R)^n.
\end{equation}
The first factor is the behaviour at the singularity, and the second factor is the behaviour at the event horizon. The transfer function is given by
\begin{equation}
    \mathcal{T} = \frac{1}{\vert b_0 \vert} \biggl| \sum_{n = 0}^{\infty} b_n \biggr|.
\end{equation}
Inserting the ansatz \eqref{eq: Leaver ansatz} in the governing Equation \eqref{eq: schrodinger} yields a five-term recurrence relation
\begin{equation}
    \alpha \, b_{n} + \beta \, b_{n-1} + \gamma \, b_{n-2} - 4 K^2 b_{n-3} + K^2 b_{n-4} = 0 \, ,
\end{equation}
with the prefactors:
\begin{align}
    \alpha &= n (n - 2 i K) \\ \nonumber
    \beta &= 1 + 7 n - 2 n^2 + L - 2 K^2 + i K(4 n - 7) \\ \nonumber
    \gamma &= 6 - 7 n + n^2 - L + 5 K^2 + i K (7 - 2 n) .
\end{align}
Additionally, $b_1$, $b_2$, $b_3$ can be expressed by $b_0$. The recurrence relation can be solved numerically, and the results are shown in Figure \ref{fig: LeaverCinf}.

\subsection{S-brane matching conditions} \label{subsec: Sbrane matching}

We have now propagated the axial metric perturbations to the S-brane and can apply the induced metric matching condition $H_1 = h_1$ and the extrinsic curvature matching condition \eqref{eq: matchingaxialvector} to determine the initial conditions for the axial perturbations to the future of the S-brane. We translate both matching conditions to the new perturbation variable $\psi$ and arrange them in a matrix $\mathcal{M}$, such that the first row is the induced metric matching condition and the second row is the extrinsic curvature matching condition. Note that we are working under the assumption that the S-brane occurs close to the singularity, i.e. $r_0 << r_S$ and thus only the largest power of $r_S/r_0$ dominates the matching. The initial conditions for the perturbation after the S-brane are therefore:
\begin{align} \label{eq: psimatching}
    \begin{bmatrix}
    \psi(t_0) \\
    \psi'(t_0)
    \end{bmatrix}
    &= \mathcal{M}^{(Bian) \ -1} \ \mathcal{M}^{(BH)} \
    \begin{bmatrix}
    \psi(r_0) \\
    \psi'(r_0)
    \end{bmatrix}  \\ \nonumber
    &= \frac{(2 L - 5)\ C_{\infty}}{2 R_0^4} \frac{b_0 a_0^{3/2}}{\det [ \mathcal{M}^{(Bian)} ]} 
    \begin{bmatrix}
    - 2 \\
    H_b \\
    \end{bmatrix},
\end{align}
where we have used the dominant behaviour of $\psi$ close to the singularity $\psi = C_{\infty} R^{-3/2}$ (see Equation \eqref{eq: regionsolutions}). The explicit forms of $\mathcal{M}^{(BH)}$ and $\mathcal{M}^{(Bian)}$ are given in Appendix \ref{appendix}, where we also argue that $\mathcal{M}^{(Bian)}$ to a good approximation depends only on $a_0$, $b_0$, $H_a$ and $H_b$ and not on $k$ and $L$.

Thus the k-dependence of the power spectrum is not changed by the S-brane, while the $L$ dependence is changed by a factor of $L - 5/2$. The power spectrum after the S-brane is still suppressed with growing $L$ but only as $L^{-2}$ instead of $L^{-4}$. 

\section{Conclusions and Discussion}

We have shown that a space-like S-brane located on a hypersurface where some curvature invariant reaches string scale can induce a non-singular transition between an external black hole space-time and an interior anisotropic cosmology. In the absence of matter in the future of the S-brane, the metric takes the form of a white hole. For matter with an equation of state close to that of a cosmological constant, space-time to the future of the S-brane is an anisotropic cosmology of long duration. In this case, our construction can be viewed as a simultaneous resolution of both the black hole and the Big Bang singularities. To obtain a cosmology compatible with current observations, a period of accelerated expansion to smooth out the initial anisotropies is required. Our setup also addresses the black hole information loss paradox since the information entering into the black hole does not get lost but goes into the new universe. More specifically, the outgoing Hawking radiation is entangled with the state inside the black hole which becomes the new universe.

We have studied the evolution of cosmological perturbations from the outside of the black hole until the onset of the cosmology phase to the future of the S-brane. We have shown that the spectral shape is modified during the approach to the S-brane. The processing of the spectrum from the black hole horizon to the S-brane can be described by a ``transfer function''. For large values of $k$, the transfer function of the power spectrum from infinity to the location of the brane scales as $k^{-1}$. For small values of $k$ where waves from the outside of the black hole have a suppressed transmission probability through the event horizon, we have computed the transfer function between a location just inside of the horizon and the location of the S-brane. Our results show that the transfer function of the power spectrum scales as $L^{-4}$, i.e. infrared modes relative to the angular directions are boosted relative to the ultraviolet modes. This is analogous to what happens in cosmological models (both inflationary and non-inflationary) where infrared modes are boosted relative to ultraviolet modes because they spend more time with wavelengths larger than the Hubble radius (see e.g. \cite{RHBrev} for a review how the spectrum of cosmological fluctuations is processed in various cosmologies). The transition through the S-brane boosts higher $L$ modes by a factor of approximately $L$. The power spectrum of the low $k$ modes after the S-brane is red with respect to $L$ and blue with respect to $k$  when compared with the initial power spectrum outside of the black hole horizon. The power spectrum at the onset of the Bianchi comsmology scales as $P(t_0) \propto P_i \ k^2 / l^{2}$.

In the case where the cosmological phase ends in a white hole, our construction can be seen as a ``Russian doll'' (Matryoshka) universe: inside the original black hole there is a cosmological phase which ends in another black hole, which then gives rise to yet another cosmological phase. Similar ideas have been explored in \cite{Slava2, Slava3}.

\section*{Acknowledgements}

We acknowledge Ad futura Slovenia for supporting J.R.'s MSc study at ETH Zurich.
The research at McGill is supported in part by funds from NSERC and from the Canada Research Chair program. R.B. is grateful for hospitality of the Institute for Theoretical Physics and the Institute for Particle Physics and Astrophysics of the ETH Zurich. LH is supported by funding from the European Research Council (ERC) under the European Unions Horizon 2020 research and innovation programme grant agreement No 801781 and by the Swiss National Science Foundation grant 179740.  

\appendix
\section{S-brane matching conditions} \label{appendix}

We here give details on the S-brane matching conditions for the axial perturbations. The matching matrices from the Equation \eqref{eq: psimatching} are:


\begin{equation}
    \mathcal{M}^{(BH)} = \begin{bmatrix}
    3 R_0^{-2} - 2 R_0^{-1} & - 2 f_0\\
    \mathcal{M}^{(BH)}_{21} & f_0^{3/2} R_0^{-1} \\
    \end{bmatrix}\, ,
\end{equation}

\begin{equation}
    \mathcal{M}^{(Bian)} = b a^{3/2} \begin{bmatrix}
    H_a - 2 H_b & 2 \\
   \mathcal{M}^{(Bian)}_{21} & - H_b \\
   \end{bmatrix} \, ,
\end{equation}
with
\begin{align}
    \mathcal{M}^{(BH)}_{21} &= \frac{3 - R_0 - 2 L R_0 + 2 (L-1) R_0^2 - 2 K^2 R_0^4}{2 \sqrt{f_0}
    R_0^4} \, ,\\ \label{eq: Mbian}
    \mathcal{M}^{(Bian)}_{21} &= \frac{1}{4}H_a^2 + \frac{7}{2} H_a H_b + H_b^2 + 2 \dot{H_b} + \frac{L}{b^2} + \frac{k^2}{a^2} \, ,
\end{align}
where $f_0 = f(r_0)$. $H_a = H_a(t_0)$, $H_b = H_b(t_0)$ and $b_0 = r_0$ are defined by the background matching conditions. 

$\mathcal{M}^{(Bian)}_{21}$ in Equation \eqref{eq: Mbian} is the only component of $\mathcal{M}^{(Bian)}$ which depends on $k$ and $L$. This dependence may be neglected as the $H_a^2$ term for example will be orders of magnitude larger for $R_0$ small. To see this, let us examine the $R_0$ dependence of terms that contribute to $\mathcal{M}^{(Bian)}_{21}$. The $k$-term: $k^2 a_0^{-2} = k^2 f_0^{-1} \sim k^2 R_0$ will be a small quantity. The $L$-term will be large: $L b^{-2} \sim L R_0^{-2}$, but from the extrinsic curvature matching conditions \eqref{eq: HubbleMatching} and the requirement that the S-brane tension be positive we have that $H_a^2 > f'^2(R_0) / 2 f_0 \sim R_0^{-3}$ is larger.


\end{document}